# A Data Source Discovery Method using Several Domain Ontologies in P2P Environments

Riad MOKADEM

*Institut de Recherche en Informatique de Toulouse (IRIT)*
*Université Paul Sabatier, Toulouse, France*
*mokadem@irit.fr*

**Abstract**

Several data source discovery methods take into account the semantic heterogeneity problems by using several Domain Ontologies (DOs). However, most of them impose a topology of mapping links between DOs. DOs and mapping links are available on Internet but with an arbitrary topology. In this paper, we propose a data source Discovery method Adapted to any Mapping links Topology (DAMT) and taking into account semantic problems. Peers using the same DO are grouped in a Virtual Organization (VO) and connected in a Distributed Hash Table (DHT). Lookups within a same VO consists in a classical search in a DHT. Regarding the inter-VO discovery process, we propose an addressing system, based on the existing mapping links between DOs, to interconnect VOs. Furthermore, we adopt a lazy maintenance in order to reduce the number of messages required to update the system due to the dynamicity of peers. The performance analysis of the proposed method shows good results for inter-VO lookup queries. Also, it confirms a significant maintenance cost reduction when peers join and leave the system.

**Biographical Statements**

Riad Mokadem received his PhD degree in Computer Science from Dauphine University in Paris, France in 2006. He is currently an Associate Professor at the University of Toulouse 3, Toulouse, France, and a member of the IRIT laboratory. His main research interests are optimization in distributed databases and large scale systems.

## 1. INTRODUCTION

In large scale query evaluation, the resource discovery step consists to look for relevant resources spread on a large scale network. It was already the object of numerous publications (Chawathe et al., 2003; Galanis et al., 2003; Huang et al., 2004; Navas et al., 2005; Talia et al., 2005; Kakarontzas et al., 2006; Marzolla et al., 2007; Paciti et al., 2007; Meshkova et al., 2008; Mokadem et al., 2012). These resources are either related to (i) computing resources, such as the discovery of a computer with an Intel Pentium T5200 with at least 2GB of free RAM or to (ii) data sources, such as a database, a XML file, a web page, a program producing data, …etc. One of the fundamental differences between both kinds of resource discovery is that for a computing resource, we look for a resource verifying certain constraints. Hence, any resource verifying the expressed constraints can be used. It is even possible, if we do not find one resource, to relax some constraints. The data source discovery researches a specific data source. Without this data source, the query evaluation is not possible. In this paper, we focus only on data source discovery in a P2P environment.

The early research works concerning resource discovery focused on keywords based discovery (Rowston and Druschel, 2001; Stoica et al., 2001; Galanis et al., 2003; Mastroianni et al., 2005; Jeanvoine and Morin, 2008; Mokadem et al., 2010]. The main principle is the search of data sources relative to a keyword. For example, we look for the data sources relatives to a keyword *Doctor*. However, the proposed methods do not take into account the heterogeneity problems of data sources. There are two types of heterogeneity: structural heterogeneity (Cruz et al., 2004) and semantic heterogeneity (Jonquet et al., 2008). The early works have focused on the consideration of the structural heterogeneity problems (e.g. difference in format, structure, complexity of data patterns). Subsequently, several studies have sought to introduce semantics in the data source discovery process. Indeed, data sources models are designed independently. This generates problems of semantic heterogeneity like synonymy (i.e. semantically similar data words are represented in different data sources) or polysemy (i.e. semantically different data are represented by the same term). For example, *Doctor* in biomedical domain is a practitioner while it corresponds to a biologist with a PhD diploma in biology. To take into account the semantic heterogeneity in the data source discovery process, three approaches have been proposed:

(a1) the approach based on the correspondence between keywords used in the data source schemas. (Reynolds and Vahdat, 2003; Rodríguez et al., 2005) proposed to publish links between keywords. Then, a discovery process with multiple keywords has to look up each keyword and return the intersection between them. The major drawback of this approach is the maintenance of links in a highly dynamic environment,

(a2) the use of a global schema or a global ontology, employed to provide a formal conceptualization of each domain (Gruber, 1995), as a pivot schema (Haase et al., 2004). However, the design of such schema or ontology is a complex task to perform in front of the huge number of data sources in large scale environments (Cruz et al., 2004) and,

(a3) the use of different domain ontologies (Navas et al., 2005). This latter approach, adopted in this paper, is the most promising because it preserves the autonomy of each domain. Hence, each application domain is associated with domain ontology (DO). Relationships links called 'mapping links' or also 'mappings' are established between these DOs in order to define correspondence links between them. In our knowledge, all methods (Nedjl et al., 2002; Halevy et al., 2003; Sartiani et al., 2004; Rodriguez et al., 2005, Akbarinia et al., 2007; Faye et al., 2007; Stevenson et al., 2013) proposed within this approach impose a particular topology on the graph formed by the ontologies and mapping links. For example: Edutella (Nedjl et al., 2002), is based on the clustering of peers that use the same schema representing a particular interest. Then, the routing protocol requires a specified HyperCup topology between super peers (Yang et al., 2003). In (Halevy et al., 2003), the topology of mapping links between ontologies must be 'two by two', i.e., in pairs. Imposing a fixed topology is a major drawback. Indeed, there are on the Internet available DOs and mapping links between them. The topology of the graph formed by these DOs and mapping links between them is an arbitrary graph. If the topology founded is not suitable for a method, some mapping links must be defined. This is a very hard task. Hence, a good challenge consists to use the existing mapping links without imposing any topology on this graph.

We have already proposed in (Ketata et al., 2010; Ketata et al., 2011a; Ketata et al., 2011c) a data source Discovery method adapted to Any Mapping links Topology (DAMT) with considering semantic heterogeneity in large scale environments. This paper extends (Ketata et al., 2011a) with mainly:

(i) a formalization of the graph formed by DOs and the existing mapping links. This formalization allows a better presentation of the DAMT method,

(ii) a proposition of an addressing system maintenance algorithm allowing a permanent access between several groups of peers using a same DO as a pivot schema and,

(iii) a more complete performance evaluation.

In DAMT method, we have grouped peers, using the same DO as a pivot schema, in a virtual organization (VO) (Iamnitchi and Foster, 2002). This allows taking into account the principle of locality (Harvey et al., 2003) that promotes the autonomy of each VO. Indeed, users of a VO often access to data sources that are in the same VO. For reasons of discovery process efficiency, the peers within the same VO are connected in a Distributed Hash Table (DHT) (Stoica et al., 2001). The data source discovery within a single VO consists in a classical lookup in a DHT. This corresponds to intra-VO queries. Concerning the inter-VOs queries, the translation of the sought concept between VOs is required. In this paper, we proposed an addressing system to interconnect VOs by exploiting the existing mapping links between DOs. This ensures a translation between different concepts in these DOs. The proposed algorithms, related to the addressing system, permit a permanent access from any VO to other in a dynamic environment. Our only hypothesis is that the mapping links graph between these DOs is connected. The DAMT method takes also into account the connection/ disconnection of peers into the system (dynamicity property of P2P environments). This requires the maintenance of the addressing system. In order to limit the excessive number of messages exchanged between peers, we adopt a lazy update of the addressing system. This permits a significant maintenance costs reduction especially in the presence of a churn effect (Mokadem et al., 2012) which occurs in the case of a continuous connection/ disconnection of peers into the system.

The rest of the paper is organized as follows: in section 2, we present the DAMT method. Mainly, we present the proposed addressing system to establish interconnections between VOs. We also discuss the maintenance of such systems while taking into account the dynamic properties of P2P environments. In the section 3, we validate the proposed method through performances comparison with three other solutions taking into account semantic heterogeneity problems. Section 4 details related work. The final section contains concluding remarks and future work.

## 2. DATA SOURCE DISCOVERY PROCESS CONSIDERING SEMANTIC HETEROGENEITY

Peer to Peer (P2P) systems are composed of a large number of autonomous peers and designed for application requiring a large amount of resource sharing (Alking et al., 2008). However, such systems are dynamic, i.e., each peer can join/ leave the system at any moment.

Data source discovery constitutes an important challenge in large scale P2P environments. It consists to search metadata describing data sources (e.g. the profile of a relation *Doctor* which is associated to a domain concept). In P2P environments, this task is complex since data sources are highly heterogeneous and constantly evolving due to data source autonomy (Ketata et al., 2011b). Furthermore, the dynamicity of peers is a major problem since the continuous joining/ leaving of peers generates prohibitive maintenance costs. In this section, we present a data source Discovery method adapted to Any Mapping Topology with considering semantic heterogeneity in large scale environments (DAMT). It also deals with the dynamicity of peers by minimizing the maintenance costs.

### 2.1 Definitions and Assumptions

To simplify the paper presentation, let consider a set of domain ontologies which form an undirected graph $G(V, E)$ with $V$ the set of vertices presenting the domain ontologies and $E$ the set of edges presenting the mapping links between these ontologies. We note that an edge exists between two vertices $v_i$ and $v_j$ in $G$ if and only if there exists a mapping link between domain ontologies $DO_i$ and $DO_j$ presenting respectively per vertices $v_i$ et $v_j$ with $i \neq j$.

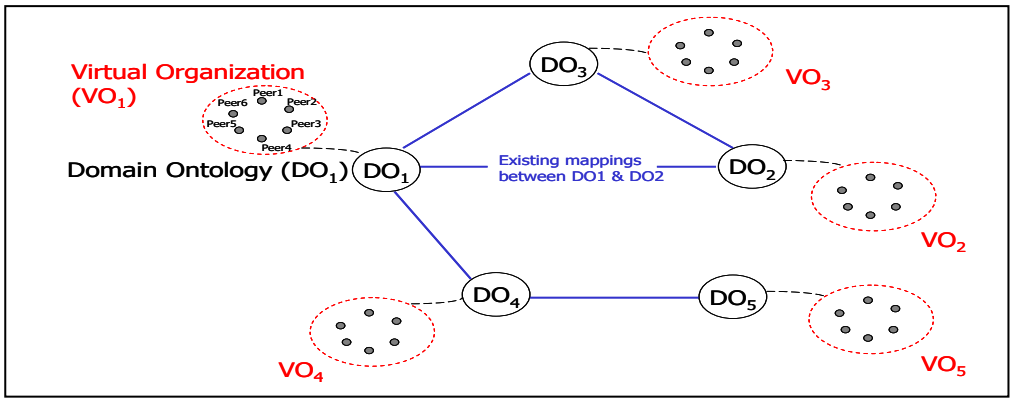

**Fig. 1.** Example of a Graph between Domain Ontologies.

For each $DO_i$ in G, we associate a virtual organization $VO_i$ as illustrated in Fig. 1. Each $VO_i$ regroups a set of peers using the same $DO_i$ as a pivot schema for managing the data sources. We affirm that one virtual organization $VO_j$ is neighbor of $VO_i$ if and only if it exists a mapping link between domain ontologies $DO_i$ and $DO_j$ used respectively per $VO_i$ and $VO_j$. We notes Neighbor ($VO_i$) a set of VOs, neighbours of $VO_i$ and connected to $VO_i$ through direct mapping links. Let also |Neighbor ($VO_i$)| be the number of VOs neighboring of $VO_i$. To ensure the completeness of data source discovery results, the graph G must be connected. In the rest of this paper, we suppose that $G(V, E)$ is connected and its topology is arbitrary. Thus, for two vertices in G there is always a path between them. This allows a translation between two DOs. Fig.1 shows an example of a connected graph of mapping links between domain ontologies $DO_1$, $DO_2$, $DO_3$, $DO_4$ and $DO_5$. As an example of a VO, $VO_1$ regroups a set of peers (Peer1, Peer2,…, Peer6) using the same domain ontology $DO_1$ as a pivot schema (Fig. 1).

After describing how VOs are connected, we present the data source discovery process of the proposed DAMT method. Using the described configuration of VOs, data source discovery queries can be classified into two types:

(i) Data source discovery within a single VO, called intra-VO discovery queries. This discovery process does not require any translation since peers used the same ontology as a pivot schema.

(ii) Data source discovery between VOs, called inter-VO discovery queries. This discovery process requires the translation of each researched concept through existing mapping links between DOs.

## 2.2 Intra-VO Data Source Discovery

The intra-VO data source discovery process consists in the discovery of data source metadata describing the researched concept in the same VO in which all data sources use the same domain ontology as a pivot schema. This solution favors the principle of locality (Harvey et al., 2003) that promotes the autonomy of each VO. In fact, a user is primarily interested in data sources that are related to its application. For example, a doctor wants to examine data sources relating to medicine. In view of its importance, we wish to (i) have an efficient mechanism for discovery process and (ii) avoid false answers (Stoica et al., 2001). This means that if the data source exists, we want to discover it.

For efficient reasons, we have proposed to associate a Distributed Hash Table (DHT) to each VO as in (Jonquet et al., 2008). Thus, peers within a same VO use the classical DHT rooting protocol (e.g., Chord protocol) (Stoica et al., 2001). Structured P2P systems using DHT have proved their efficiency with respect to the scalability and research process. In addition, they have the characteristic to avoid false answers. Recall that the complexity to find a peer responsible of a data source is $O(\log(N))$ where $N$ is the number of peers (Stoica et al., 2001).

## 2.3 Inter-VO Data Source Discovery

Inter-VO data source discovery process consists to search a concept in a set of VOs which are associated to different DOs. Thus, an inter-VO data source discovery query Q providing from a peer $\epsilon$ $VO_i$ consists to look for metadata of data sources available in $VO_j$ with $i \neq j$. In other terms, metadata discovery queries must be propagated between VOs. The existing mapping links between DOs are exploited. For this aim, we have proposed an addressing system which assures a permanent access from any VO to other in a dynamic environment. This is done in order to propagate the data source discovery queries. The following sub-sections describe the proposed addressing system and the inter-VO data source discovery process.

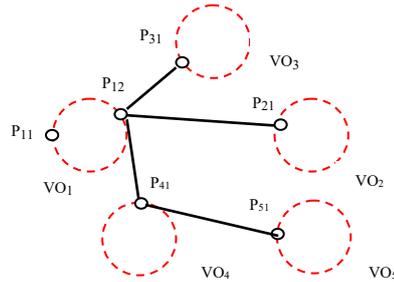

**Fig. 2.** Example of Interconnection between VOs through Access Points.

### 2.3.1 Addressing System

This section describes the addressing system allowing communication between VOs. In order to interconnect peers of different VOs and then permit a communication between a peer $\in VO_i$ and peers in neighbor($VO_i$), we associate for each peer $P_i \in VO_i$ |Neighbor ($VO_i$)| access points. Let $AP_i$ the access point set of $P_i$. Each access point $P_j \in AP_i$ is one peer of $VO_j \in$ Neighbor ($VO_i$). Hence when a peer $P_i$ wants to propagate a data source discovery query to these neighbor VOs, access points and exiting mapping links between $DO_i$ and $DO_j$ are used. In order to avoid that a peer forms a bottleneck or constitutes a single point of failure, we ensure that several access points $P_j$ of a peer $P_i \in VO_i$ reference different peers in $VO_j \in$ Neighbor ($VO_i$). Fig. 2 illustrates examples of access points: the bold lines show mapping links between VOs. For example, $P_{12} \in VO_1$ can communicate with peers of $VO_2$ thanks to its access point $P_{21}$.

**2.3.2 Inter-VO Data Source Discovery Process**

Suppose that a given peer $P_i$ submits a discovery query $Q$. The inter-VO data source discovery process consists to propagate $Q$ towards others VOs through the access points $\in AP_i$. Fig. 3 shows the inter-VO discovery algorithm which describes the data source discovery process. When $P_i$ contacts its access point, the researched concept C is translated through the existing mapping rules between DOs. Every peer receiving a discovery query: (i) execute an intra-VO data source discovery query and (ii) propagate the query to all its access points. Hence, a lookup function, noted *Lookup( )*, is evaluated for each each $P_j \in AP_i$ in order to search the concept C in $VO_j \in$ Neighbor ($VO_i$). To avoid an endless propagation of a discovery query, we define a Time to Live (TTL) which corresponds to the maximal path length in G that a discovery query follows along the discovery process, i.e., it constitutes the maximal number of propagation hops in G that a discovery query can follows. It is initialed at the graph diameter.

```
Inter-VO Discovery (!C: the researched concept,
                    !Pi: the peer ∈ VOi , initiator of the discovery query,
                    ?M: metadata describing data sources, to be discover,
                    ?Path: resource discovery path)
        {
        Path=∅:
        M ← Lookup(C, VOi, Pi, Path);
                            //Intra-VO data source search
        TTL ← TTL – 1;
        If(TTL != 0) then
          For each Pj ∈ APi
            M ← M U Lookup(Translate(C,VOi,VOPj),VOj, Pi, Path U VOj);
                            //Inter-VO data source search
          If(not Empty(M))
            then Return(M, Path U VOj);
        }
```

**Fig. 3.** Inter-VOs Data Source Discovery Algorithm.

In a P2P environment, each peer can joins or leaves the system at each moment. Hence, access points $P_j \in VO_j$ of a peer $P_i \in VO_i$ could have left the system. To avoid the endless waiting of an access point response, we define an interval of time noted RTT (Round-Trip-Time). Then, if $P_j$ does not respond to $P_i$ after a RTT, it is considered to be disconnected. In consequence, the peer $P_i$ contacts the nearest neighbor $P_{i+1}$ in its VO in order to request its access point toward $VO_j$. Then, the inter-VO data source discovery query is propagated to $VO_j$. If this access point responds, $P_i$ takes also the opportunity to ask the nearest neighbor of the founded access point $P_{j+1}$ in $VO_j$. This neighbor will become the new access point of $P_i$ to $VO_j$. This allows a $VO_i$ to have several different peers as access points to another $VO_j$. Thus, we avoid that a single peer forms a bottleneck. If all neighbors of $P_i$ have been contacted without any response, the concerned $VO_j$ is considered to be disconnected and only the administrator can reconnect it. At the end of the discovery process, the response of data source discovery is sent to $P_i$, the peer initiator of the data source discovery query. This response contains: (i) metadata, noted M in Fig. 3, describing the discovered data sources and (ii) the path, noted *Path* in Fig. 3, constituted of a sequence of edges representing the mapping links that the discovery query followed along the discovery process in order to translate the response. This path will be used for the response routing to $P_i$.

**2.4 System Maintenance**

The leaving and joining of peers is very common in P2P systems (dynamicity property of P2P environments). In consequence, the maintenance of the system is required. We distinguish two types of maintenance in our system: (i) maintenance of the DHT and (ii) maintenance of the defined addressing system that impacts the discovery process. We will not detail the first case since the system maintenance is done by a classical maintenance of a DHT (Stoica et al., 2001). Recall that in structured

P2P Chord systems used in our system, the connection/ disconnection of one peer generates $Log^2(N)$ messages when *N* is the total number of peers (Stoica et al., 2001). In this section, we interest only on the addressing system maintenance. Hence, maintaining such system requires the updating of all access points. In other terms, the maintenance of the addressing system consists in defining how the access points are updated. Recall that the DAMT method is efficient when a VO is composed by more than one peer. The fact that one VO is composed by a single peer is a trivial case when a peer leaves the system. In this case, the peer leaving requires the administrator intervention in order to later reconnect the corresponding VO to other VOs. Throughout this section, we describe both the connection and disconnection steps.

**2.4.1 Peer Connection Process**

When a peer joins the system, all its access points must be defined. We based on the same technique used when an access point is not available during the inter-VO discovery. Fig. 4 shows the access points' definition algorithm. Suppose that a new peer, called *NewP*, is connected to a $VO_i$. It needs access points $P_j$ to access all other $VO_j \in Neighbor(VO_i)$. For this, it contacts, via the *NeighborPeer( )* function, its nearest neighbour in $VO_i$ (noted *Following* in Fig. 4) to get its access points to $Neighbor(VO_i)$. Then, *NewP* contacts all these peers $P_j$ by requesting their nearest neighbor in each $VO_j \in Neighbor(VO_i)$. These neighbors will become the new access points of *NewP* in each $VO_j \in Neighbor(VO_i)$. In other terms, if an access point in $VO_j$ responds (*Found=True*), its neighbor is asked. This later will constitutes the new access point for *NewP* to this $VO_j$.

As described the inter-VO discovery process, if an access point towards a $VO_j$ ($j \neq i$) does not respond during a certain period of time noted RTT (Round-Trip-Time), it is considered to be disconnected. Then, *NewP* contacts the neighbor of its neighbor. It repeats this process recursively until getting a connected peer in $VO_j$. The test is done via the *Check( )* function.

```
AccessPoint_Definition(!NewP: a new peer joining the system)
  {
    For j ∈ Neighbor(VOᵢ)
    { Found = False;
      Following = NeighborPeer (NewP);
      While(Following != NewP) and (not Found)
      {
        Found = Check(Following → Pⱼ);
        If(Found) then   Pⱼ → NeighborPeer (Pⱼ);
         Else           Following = NeighborPeer (Following);}
    }
  }
```

**Fig. 4.** Access Points' Definition Algorithm.

**2.4.2 Peer Disconnection Process**

A peer disconnection can be divided into *friendly leaves* and *peer failures* (Meshkova et al., 2008). Friendly leaves enable a peer to notify its overlay neighbors to restructure the underlying topology. Peer failures possibility is more complex and seriously damages the structure of the overlay with data loss consequences. In this section, we first explain the maintenance process in case of friendly leaves and then we discuss about peer failure.

In order to explain, the maintenance process in case of friendly leave, suppose that one peer, called $P_{Disc} \in VO_i$ disconnects from the system. The first step is to maintain the DHT system. This is a classical DHT maintenance (Stoica et al., 2001). However, the peer $P_{Disc}$ can be an access point for a peer belonging to another $VO_j$ (with $i \neq j$). Hence, the addressing system must be updated. Two solutions emerge. In a first solution, $P_{Disc}$ can propagate the information to every $VO_j$ towards which it is connected. This strategy proceeds by flooding in VOs. We do not adopt it since the flooding generates large volume of unnecessary traffic in the network. Other solution, adopted in our system, consists to apply a lazy maintenance. None of the $VO_j$ (with $i \neq j$) is informed by this disconnection. The access points towards this $VO_j$ will be updated during an inter-VO data source discovery process. Indeed, during the inter-VO data source discovery process, the opportunity is taken to update all access points.

This strategy reduces the number of peer messages. The system is updated every time the data source discovery process is performed. Hence, as data source discovery process is frequently performed as the system is up-to-date.

## 3. PERFORMANCE EVALUATION

In order to validate the proposed DAMT method, we evaluate it by comparing its performances to those of three other data source discovery methods taking into account the semantic heterogeneity of data sources. These methods are: (i) data source Discovery according to the Super-Peer topology Method (DSP) as in (Faye et al., 2007), (ii) data source Discovery method that we call Two by Two peers (D2b2) as in (Halevy et al., 2003) and (iii) data source Discovery by Flooding Method (DFM) as in (Chawathe et al., 2003). In this evaluation, we respect topologies imposed in each of implemented methods. Thus, in the DSP method, the imposed mapping topology is based on super peer model. Each VO has a single access point to another VO and super-peers are interconnected randomly with a minimum of four neighbors in other VOs. Regarding the method D2B2, the mapping topology imposed is the connection 'two by two'. In the DFlooding method the discovery process relies on a flooding mechanism. In following experiments, we have not experiment with intra-VO discovery queries since the discovery process in this case corresponds to a classical DHT lookup. Throughout this section, we deal with two classes of experiments:

(i) The inter-VO lookup performances by evaluating the data source discovery response time. The response time is the time elapsed between transmitting the data source discovery query and receiving the corresponding response.
(ii) The impact of the peers' connection/ disconnection with respect to the maintenance of underlying system.

### 3.1 Simulation Environment

To evaluate performances of the proposed DAMT data source discovery solution, the NS2 simulator (Issariyakul and Hossain, 2008) is adopted to simulate the inter-VO data sources discovery according to topologies of the four compared methods. A virtual simulated network of ten thousand (*10000*) homogeneous peers is simulated with homogenous bandwidth networks *100 Mb/s*. We suppose that the full communication time (hop cost) between two VOs is fixed to 10 ms. We based on Open Chord (Stoica et al., 2001), one implementation of the Chord DHT, to simulate the data source discovery in the DHTs. Recall that we used a simulation environment since we have not this number of peers nor the network infrastructure.

### 3.2 Inter-VO Data Source Discovery

We measure the response time of the inter-VOs data source discovery process for the four compared methods. In first experiments, we have experimented with a single data source discovery query. We have varied the number of VOs. But, the total number of peers in the system stays always constant. Then, other experiments show the impact of the data source discovery query number transmitted per second and per peer on the response times in order to measure the capability of each method to be scalable or not.

#### 3.2.1 Single Data Source Discovery Query Performance

In this sub section, we experiment with only one data source discovery query. Fig. 5 shows the evolution of data source discovery response time when the number of VO is varied from 2 to 100. These two end values correspond to a configuration with 5000 peers and 100 peers per VO respectively. Fig. 5 shows that the response time of the D2B2 method is the largest compared to the other three methods. This is due to the length of the longest path traversed to discover data sources within this method. We have also almost similar response times between DSP, DFlooding and DAMT methods when the VO number does not exceed 10. However, DAMT has the advantage of not depending of the peer availability as in DSP method. When the number of VOs exceeds ten (10) VOs, DFlooding and DAMT methods show better results in terms of response time. They have almost similar response times with a small advantage to the DFlooding method. However, the graph of mapping links between ontologies in the DFlooding method must be a complete graph which requires intensive intervention of the administrator. This is not the case in the DAMT method in which the only constraint is to have a connected graph.

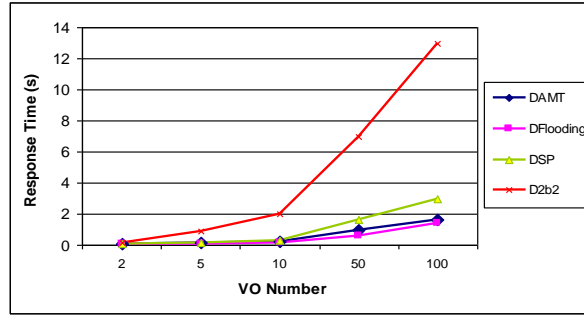

**Fig. 5.** Response Times when using a Single Discovery Query.

### 3.2.2 Impact of the Data Source Discovery Query number on the Response Time

In these experiments, we measure the impact of simultaneous data source discovery queries per second and per peer on the data source discovery response time. This provides information on the ability of each method to be scalable in the presence of high number of discovery queries. We have varied the number of data source discovery queries submitted to each peer within a VO. Fig. 6 shows response times when the number of VOs is equal to 10 VOs (Fig. 6- left) and 100 VOs (Fig. 6- right). These configurations correspond to systems with 1000 peers and 100 peers respectively in each VO.

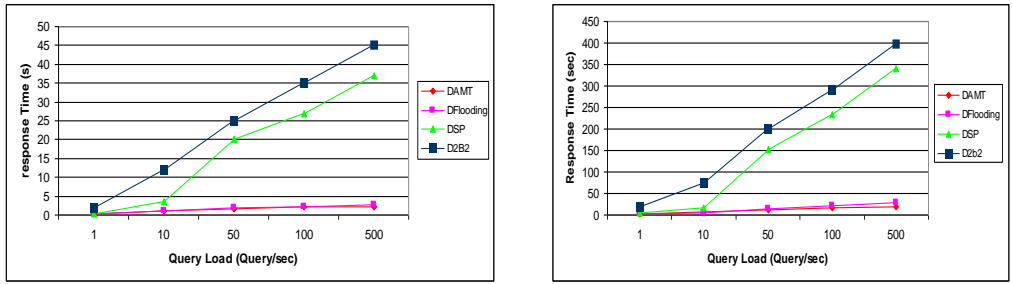

**Fig. 6.** Impact of the Query Load on the Response Time with 10 VOs (left) and 100 VOs (right).

We observe that results obtained by DSP and D2b2 methods are much higher than DFlooding and DAMT results as shown in Fig. 6- left. This is due to the saturation of peers causing bottlenecks in the DSP method and daisy chain in the D2B2 method. For the DSP method, this is valid especially when the simultaneous messages exceed 10 per second. We can also observe that the best results were obtained by DFlooding and DAMT methods which have almost equivalent results. Our response times are slowly greater than DFlooding response times when we experiment with less than 20 queries per second. However, slightly better results are observed with DAMT method when we experiment with more than 50 simultaneous queries per second. We get a very low growth rates (only 0.6 sec for 40 additional queries per second). Concerning the experiments with 100 VOs, results obtained by DSP and D2b2 methods are also much higher than DFlooding and DAMT results. Fig. 6-right shows almost similar results between DFlooding and DAMT methods when the number of simultaneous queries is less than 25 simultaneous queries per second. From this value, DAMT response times are slowly better than DFlooding response times. This is due to the fact that multiple discovery queries in our method may require the intervention of several different access points when these queries generate some bottleneck at some peers in the DFlooding method.

The DAMT method substantially improves the response time obtained by D2b2 and DSP methods. When we experiment with 10 simultaneous queries per second, we observe that the response times generated by our method are 3.5 (10 respectively) times smaller than the response times generated by the DSP method when we experiment with 10 VOs (100 VOs respectively). From 10 queries per second, the save time is most important (Fig. 7).

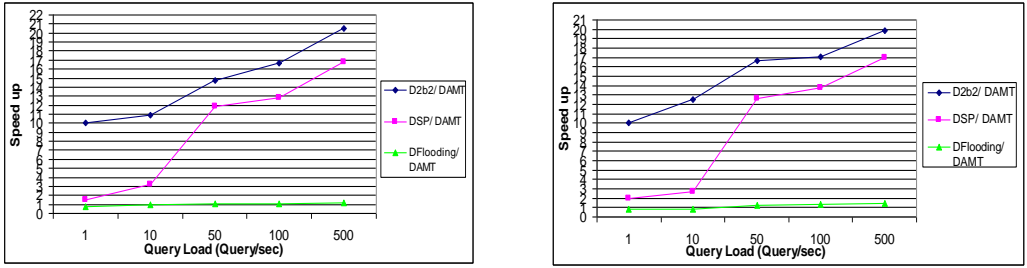

**Fig. 7.** Speed up (Response Times) with 10 VOs (left) and 100 VOs (right).

When we experiment with 500 simultaneous discovery queries, we can observe that the response time generated by DAMT method is 17 (20 respectively) times smaller than the average response time generated by the DSP (D2b2 respectively) method. Compared with the DFlooding method, our method produces slowly poorer performance when the number of resource discovery queries per second is reduced (less than 20 queries per second). DFlooding response times are 20% (18% respectively) less than DAMT response times when we experiment with 5 simultaneous queries per second with 10 VOs (100 VOs respectively) configuration. A save time, whatever small, is obtained from 25 discovery queries per second. DAMT response times are 10% less than DFlooding response times when we experiment with 500 simultaneous queries per second with 10 VOs configuration. This is also due to the fact that we have bottleneck at some peers in the DFlooding method when several different access points are used in the DAMT method. Thus, it seems more reasonable to have important simultaneous queries in a large scale environment composed with more than 10000 peers.

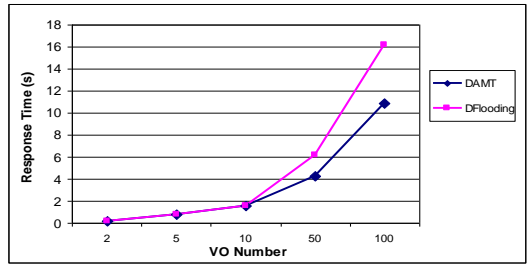

**Fig. 8.** Impact of the VO Number on the Response Time with Simultaneous Queries per Second.

In order to compare performances of DAMT and DFlooding methods with a better precision, Fig. 8 shows the evolution of the response time of the two methods when we experiment with 20 simultaneous queries per second and per peer when varying the number of VOs between 2 to 100. We have chosen to experiment with this number of simultaneous resource discovery queries since we have obtained similar response times for DAMT and DFlooding methods when we experiment with 20 queries per second and per peer. We have a small advantage to the DFlooding method when we experiment with only two (2) VOs and almost similar response times when the VO number is between 2 and 10. When this number exceeds ten (10) VOs, we have an advantage to the DAMT method. In fact, DFlooding method is based on intensive flooding that generated bottleneck in some peers which affect performances. The save time is more important (40%) when we experiment with 100 VOs. Hence, it seems more reasonable to have more than ten (10) VOs in a configuration with ten thousand peers.

### 3.3 System Maintenance

We measure the impact of the joining/ leaving peers on the system maintenance with respect on the number of messages required to maintain the system. We simulate that several peers join and leave the system but the total number of peers stays appreciatively constant. We opted for a configuration composed of 100 VOs. In other experiments, we studied the impact of access peer session's length on the system maintenance.

### 3.3.1 Impact of Joining/ Leaving of Peers on the System Maintenance

We evaluate the maintenance costs by measuring the number of messages required to maintain the system. It is clear that maintaining a DHT generates greatest costs especially when several peers join/leave the system. But, this is valuable for all the compared solutions.

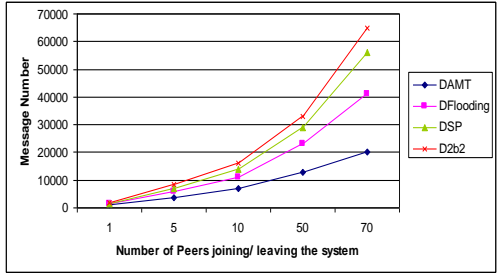

**Fig. 9.** Impact of the Number of Connected/ Disconnected Peers on the System Maintenance.

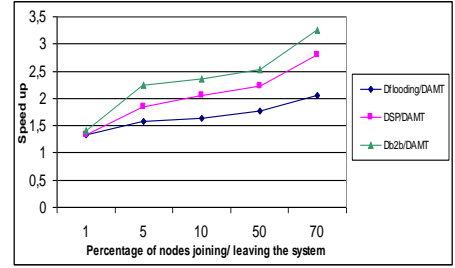

**Fig. 10.** Speed up (Number of required messages to update the system).

Fig. 9 shows the number of the required messages to maintain the system when the number of VOs is equal to ten (10) VOs. The gap between the four curves grows especially when the number of peers joining/ leaving the system is increased. The number of messages needed to maintain the system with the DSP and D2b2 methods is the higher. Indeed, if a super peer leaves or arrives in the system, all the 'leafs' peers should be updated by using the DSP method. The D2b2 method generates the most important maintenance cost. This is due to the topology used. Better results are observed for the DFlooding method which requires less than 11000 for 10 connected/ disconnected peers when the maintenance reaches 14000 messages in the two other methods for the same number of connected/ disconnected peers. When we experiment with 20 VOs, the number of these messages exceeds 20000 and 18000 for the DSP and DFlooding methods respectively. Our method requires fewer messages for updating the system that all other methods. Thus, only 7200 messages (respectively 11000 messages) are required to maintain the system when 10 peers join/ leave the system in a configuration with 10 VOs (respectively 20 VOs). In fact, most of messages in the DAMT method are essentially those required to update the DHT. The use of a lazy maintenance in our solution allows significant reduction in the number of these messages needed for access point update unlike other methods. Thus, access points of a peer referencing peers which have leaving the system are updated only when the inter-VO discovery process occurs in the DAMT method while all peers must be contacted to update their access points on each VO in the DFlooding method.

We have also interest to the impact of the percentage of the peers connection/ disconnection on the number of messages required to update the system. The percentage concerns the number of peers which join/ leave the system comparing to the total number of peers in the system. In these experiments, each VO is composed of 100 peers. We simulate the percentage of peers which join/ leave the system. Fig. 10 shows the saves obtained by the using of the DAMT method comparing to the three other methods. This save is about 30% comparing to the three compared methods when the connection/ disconnection concerns only one peer. When this percentage increase, the save is more important especially when we compare the DAMT to the Db2b method. When more than 50% of peers in the system join/ leave the system, the gap increases even more. We have a save of 300% (resp. 200%) comparing to the Db2b (resp. Flooding) method when we deal with a disconnection of 70 peers. This is explained by the fact that a disconnection of peers in DAMT method requires only the DHT update when it requires several communications between peers in the other compared methods.

**3.3.2 Impact of the Access Peer Session's Length on the System Maintenance**

We also compare the four methods by taking into account the mean session duration of peers. We define a session length in our system as follow: we focus only on an observation interval of fixed duration T (equal to 10 sec). Then, each session length is represented as a percentage of T and corresponds to the duration in which the peer stays in the system, and then leaves it. It corresponds to values {10%, 20%… 100%} in Fig. 11. Let $R$ the remaining time which corresponds to T minus the session length. Simultaneous inter-VO resource discovery queries (100 messages/ sec) are sent at the beginning of T. Then, disconnections of 5% of peers occur in each VO. The concerned peers are soon replaced by other peers. The result collection is done when the duration T is finished. In other terms, when T is finished, we measure the number of messages required to finish both maintenance and discovery processes. Fig. 11 shows the number of these messages when we vary the session length. In these experiments, we have fixed the number of VOs to 10 and deal with 20 simultaneous discovery queries.

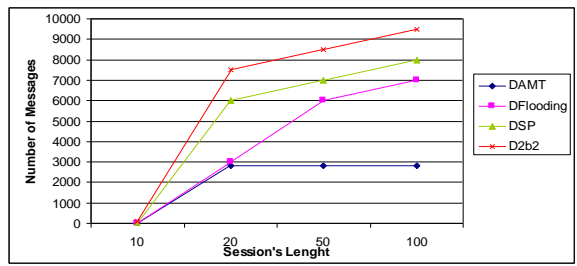

**Fig. 11.** Impact of the Total Session's Length on the Number of Update Messages.

When the session length represents only 20% of T, DFlooding and DAMT performances are almost similar. The disconnection of some peers generates an update of the system which is done in the remaining time (80%) of T in the DFlooding method. The disconnection of several access points will not generate any update in the DAMT method since this update is done during the discovery process. When we measure the number of messages after T period, the update process of DFlooding is finish which explains the similarity between the two curves. Compared to the two other methods, DFlooding and DAMT curves are better than DSP and D2b2 curves. It is due to the fact that updating the system in DSP and D2b2 methods requires a time much greater than *R*. When the session length exceeds 20% of T, as the session length is important as the gap between DAMT and DFlooding curves increase. It is due the fact that several peers have not finished the update process in the DFlooding method when T is finished. This process concern only peers concerned by the discovery process in the DAMT method.

### 3.4 Discussion

Performance evaluation has allowed us to compare our method to three other data source discovery methods (D2b2, DSP and DFlooding methods) taking into account semantic heterogeneity problems. In all these methods, intra-VOs lookups are based in a classical DHT lookup. This makes the results equivalent for this type of discovery. For the inter-VOs discovery process, the performance evaluation of our method shows significant response time reduction compared to D2b2 and DSP methods. The gain is significantly important when we deal with several simultaneous discovery queries per second. This gain also increases when systems have a greater number of VOs, which is often the case in large environments. Compared with the DFlooding method based on a flooding process, this later has slightly better response times (21%) in the case of a single discovery query. When the number of simultaneous discovery queries increases, our method has better performance than the DFlooding method. It seems more reasonable to have more than 20 queries per second in a large scale environment than having only one query per second. Regarding the system maintenance, evaluation of our method shows very good results. Due to the lazy maintenance adopted in the DAMT method, system maintenance costs are significantly reduced compared to the three methods cited above.

## 4. RELATED WORK

A lot of resource discovery methods based on P2P techniques (Iamnitchi et al., 2002, Talia et al., 2005; Marzolla et al., 2007; Samad et al., 2009; Mokadem et al., 2010; Mokadem and Hameurlain, 2011) do not deal with the problem of semantic heterogeneity. They are limited to keywords search which may returns erroneous results because of the strongly semantic heterogeneity when data sources are developed autonomously. Dealing with this problem, (Molto et al., 2008; Comito et al., 2009; Tao et al., 2010, Lopes and Botelho, 2012) have focused on the semantic by using the web service discovery. But, such systems typically employ flooding and random walk to locate data which results in much network traffic. In order to provide scalability and efficient query routing, some research works adopted DHT as alternative to centralized or hierarchical approaches. (Pirro et al., 2008) proposed to combine DHT and web services through the ERGOT system. Semantic closed services are clustered to support semantic search. However, these methods generate high maintenance costs in dynamic environments.

Dealing with these problems, some research works proposed to establish a correspondence between keywords used in the schemas of data sources. (Halevy et al., 2003) integrates sources that are semi-structured data based on XML, data model and XQuery language. Then, peers interested by exchanging data establish semantic links 'mapping' between them in 'two per two' way. The main insufficiency of this method is the difficulty of describing these semantic links and maintaining them. In Hyperion (Arenas et al., 2003), a conventional database management system (DBMS) augmented with a P2P interoperability layer, the resource research has focused on the management of metadata that enables data sharing and coordination between the peers. Mapping tables (Kementsietsidis et al., 2003) are used to achieve data integration between peers. However, many efforts are needed to create these tables in a semi-automatic manner and to maintain these tables up-to-date. In XPeer (Sartiani et al., 2004), peers export a tree shaped data guide description of their data, which is automatically inferred by a tree search algorithm. Each query is translated into an algebraic expression and sent to the super peer network. Then, comparisons of algebraic expressions permit to respond to the initial query. PeerDB (Siong et al., 2005) is reportedly the first Peer Data Management Systems (PDMS) implementation taking into account the semantic properties. It used keywords for each relation and attribute. To process queries, it exploits mobile agents for flooding the query to the peers. (Kantere et al., 2009) focuses on the successive translations of queries. Then, peers update their knowledge according to the accuracy of the answers they receive. The main limitation of these routing approaches is the using of unavoidable flooding with limited semantic information.

To tackle these problems, some methods proposed the using of ontologies, employed to provide a formal conceptualization of each domain. Some studies were based on a global ontology/ schema shared by all peers (Alking et al., 2008; Akbarinia and Martins, 2007; Cruz et al., 2007; Haase et al., 2004; Huebsch et al., 2005). PIER (Huebsch et al., 2005), a structured P2P system constitutes a good example in which all peers share a standard schema. In Bibster (Haase et al., 2004) also, queries are formulated according to a shared common ontology. The query is first routed according to the expertise descriptions known by each peer. The flooding is used only if there is an unsuccessful routing of the query. However, it is still a complex task to design a global schema, due to the strong diversification of domains in dynamic large scale environments. Other works use several schemas (Heine et al., 2004; Juan, 2010). In (Heine et al., 2004), each peer has its own ontology represented as a classification Directed Acyclic Graph (DAG), which captures subsumption relation between concepts in each ontology. Although that peer's ontology is not complete, it can be completed by ontologies of other peers. Then, the discovery process is based on a DHT algorithm that distributes local DAGS among peers of the P2P network. A distributed virtual view of all peers' graphs is conceived. Thus, the discovery process consists to determine the peer which store information for this concept including a list of super concepts according to the DAG. METEOR-S (Verma et al., 2005) is suggested for adding semantics to the Web service standard by providing algorithms that annotate WSDL files with relevant ontologies. It uses an ontology-based approach to organize multiple registries into domains. When a peer sends a query to a peer group, it will receive registries ontology and need to choose the most relevant domain specific ontology. This approach demands high performance especially for the client group leader which has the responsibility of selecting relevant domain for each peer. In SenPeer (Faye et al., 2007), peers are connected to super peers according to their semantic domain. However, super peers generate bottleneck and fault tolerance problems. (Tao et al., 2010) propose to combine a scalable DHT with an ontology based Information Service (DIS) for a grid system by organizing resources into a DHT ring. However, this solution does not take into account the dynamicity of nodes since the VO manager consider only stable VOs. (Souza et al., 2011) proposes to computes a cluster ontology summary for peers using the same schema. Then, each incoming peer searches for the closest cluster. However, this solution requires the explicit representation of clusters. The Gossiping process introduced by (Kermarrec and Van Steen, 2007) also deals with semantic heterogeneity problems. (Cerqueus et al., 2012) used a Gossiping process to share the semantic links (e.g., correspondences) across the system. Indeed, if the peers know a lot of correspondences between entities of different ontologies, semantic heterogeneity problems are less frequent. Another work (Amato et al., 2012) deals with resource discovery through a mix of logical formulas. Data are linked together and enriched with additional knowledge coming from the external source of information. This knowledge may be exploited during the matching process for preferring a rule rather than another one. Although this work gives promising theoretical results, authors have not experiment the proposed method in a large scale environment.

(Cai and Frank, 2004; Kokkinidis and Christophides, 2004; Sidirourgos and Kokkinidis, 2005; Kaoudi et al., 2007) were based on RDF schemas. RDF was chosen as it is semantically rich and supports extensive querying capabilities without resorting to network flooding. It is used to represent both data sources and queries. RDFPeer (Cai and Frank, 2004) indexes each RDF triple to support semantic RDF query. Then, a query with multiple keywords is resolved by using the DHT which lookup each keyword. The final result is the intersection of all intermediate results. Authors in (Kokkinidis and Christophides, 2005) describe middleware based on P2P system for evaluating queries in the RDF Query Language (RQL) using RDF Schema knowledge. They focus on the construction and optimization of query plans. Instead of using DHTs, they use so called semantic overlay networks, which are groups of peers sharing the same schema. However, most of the proposed methods exploit a specific imposed topology between domain ontologies by organizing peers into semantic communities. They are typically dependent on this topology (i.e. unstructured, structured or hybrid). This requires an intensive intervention of the administrator when some mapping link does not exist.

## 5. CONCLUSION

We have proposed a data source discovery method taking into account both semantic heterogeneity and dynamicity of peers in an unstable large scale P2P environment. We group all peers using the same domain ontology (DO) in a virtual organization (VO) in order to favor the locality principle. Within a VO, the data source discovery process (intra-VO discovery) is based on a lookup in a classical DHT. DHTs showed their efficiency and have the advantage to avoid false answers. Regarding the inter-VO discovery process, we defined an addressing system which exploits the mapping links set existing between the various DOs. The originality of our method is to not impose any topology on the graph formed by DOs and mapping links between them. Furthermore, our discovery method allows a permanent access between VOs in a dynamic environment. In addition, we adopt a lazy maintenance in order to decrease the update cost generated by the continuous joining/ leaving of peers.

We have compared performance of our method to those obtained by three other data source discovery methods taking into account the semantic aspect. With regard to both DSP and D2b2 methods, using as us mapping links between DOs, the performance evaluation showed a significant improvement of the response time for the inter-VO data source discovery queries. We also compared our method with another method based on the flooding principle (DFlooding method). While sacrificing response times for a reduced number of queries per second, we have an improvement with a more important number of simultaneous queries per second. Hence, it seems more reasonable to have more than 20 queries per second in a large scale environment than having only one query per second. In addition, the proposed DAMT method is adapted to any mapping topology unlike other compared discovery methods. By imposing a particular topology on the graph formed by ontologies and mapping links, most of methods in the literature ask the definition of mapping links if the graph founded (e.g., on Internet) is not suitable. The definition of these mapping links is always a complex and hard task since it demands the knowledge of both ontologies. Finally, our method obtains a significant reduction of the system maintenance cost generated by the joining and leaving of peers. This reduction is due to a lazy maintenance of the proposed addressing system. Our method can be useful in large scale environments since it generates less traffic network. Further work includes more performance studies especially in a real platform with a high number of nodes.